%% file: ir-nnlm-arxiv.tex
\documentclass[english]{article}
\usepackage[T1]{fontenc}
\usepackage[utf8]{inputenc}
\usepackage{url}

\usepackage{amsmath}
\usepackage{amsfonts}
\usepackage{fullpage}
\author{
    N. Despres\thanks{Sorbonne Universités, UPMC Univ Paris 06, CNRS, LIP6 UMR 7606, 4 place Jussieu 75005 Paris}\\
    {\small nicolas.despres@gmail.com}
    \and
    S. Lamprier\footnotemark[1]\\
    {\small sylvain.lamprier@lip6.fr}
    \and
    B. Piwowarski\footnotemark[1]\\
    {\small benjamin@bpiwowar.net}
}

\usepackage[disable]{todonotes}

\makeatletter

\usepackage{tikz}
\usetikzlibrary{positioning, shapes}
\makeatother
\date{}

\usepackage{babel}

\begin{document}

\title{Parameterized Neural Network Language Models for Information Retrieval}
\maketitle
\begin{abstract}
Information Retrieval (IR) models need to deal with two difficult issues, vocabulary mismatch and term dependencies.
Vocabulary mismatch corresponds to the difficulty of retrieving relevant documents that do not contain exact query terms but semantically related terms.
Term dependencies refers to the need of considering the relationship between the words of the query when estimating the relevance of a document.
A multitude of solutions has been proposed to solve each of these two problems, but no principled model solve both.
In parallel, in the last few years, language models based on neural networks have been
used to cope with complex natural language processing tasks like emotion
and paraphrase detection. Although they present good abilities to cope with both term dependencies and vocabulary mismatch problems,
thanks to the distributed representation of words they are based upon, such models could not be used readily in IR, where
the estimation of one language model per document (or query) is required. This is both computationally unfeasible and prone
to over-fitting.
Based on a recent work that proposed to learn a generic language model that can be modified through a set of document-specific parameters, we explore use of
new neural network models that are adapted to ad-hoc IR tasks.
Within the language model IR framework, we propose and study the use of a generic language model as well as a document-specific language model. Both can be used as a smoothing component, but the latter is more adapted to the document at hand and has the potential of being used as a full document language model.
%where individual two    A recent work however proposed a neural network language model where to performed document-dependent modifications of a %modify the probability distribution of a
 %generic language model, % by using a single vector,
 %thereby allowing to develop language models for IR that are able to
%cope with both term dependencies and vocabulary mismatch problems.
We experiment with such models and analyze their results on TREC-1 to 8 datasets.
\end{abstract}
\newcommand{\argmax}{\mathop{\textrm{argmax}}}
\newcommand{\real}{\mathbb{R}}
\newcommand{\note}[2][]{\todo[inline,author=#1]{#2}}

\listoftodos

\section{Introduction}

To improve search effectiveness, Information Retrieval (IR) have sought
for a long time to properly take into account term dependencies and
tackle term mismatch issues. Both problems have been tackled by various models, ranging from empirical
to more principled approaches, but no principled approach for
both problems have been proposed so far. This paper proposes an approach based on recent
developments of neural network language models.

Taking into account dependent query terms (as compound words for instance) in document relevance estimations usually increases the precision of the search process. %by considering documents
%in which inter-dependent query terms (as compound words for instance) are
This corresponds to developing approaches for identifying term dependencies and considering spatial proximity of such identified linked terms in the documents. %close to each other.
Among the first proposals to cope with term dependency issues, Fagan et al.  \cite{Fagan1987PhraseIndexing}
proposed to consider pairs of successive terms (bi-grams) in vector space models. The same
principle can be found in language models such as in \cite{Song1999GLM} that
 performs mixtures of uni- and bi-gram models. Other works have sought to combine
the scores of models by taking into account different co-occurrence patterns, such as \cite{Metzler2005MRF} which proposes a Markov random field model to capture the term dependencies of queries and documents. %, involving a various number of query terms.
In each case, the problem comes down to computing accurate estimates of n-gram language models (or variants thereof), i.e. language models where the probability distribution of a term depends on a finite sequence of previous terms.

On the other hand, taking into account semantic relationships (such as synonymy) increases recall by enabling the retrieval of relevant documents that do not contain the exact query terms but use some semantically related terms.
%containing the query terms, which
This is particularly important because of the asymmetry
between documents and queries.
Two main different approaches are used to cope with this problem.
The first is to use (pseudo) relevance feedback to add query terms that were not present in the original query.
The second is to use distributed models, like latent semantic indexing,
where terms and documents are represented in a latent space, which might
be probabilistic or vectorial.
None of these approaches have taken term dependencies into account.

In this paper, we show that neural network language models, that leverage a distributed representation of words, handle naturally terms dependence, and have hence an interesting potential in IR.
Neural Language Models \cite{Bengio2003NNLM} have been successfully
used in many natural language processing tasks, like part-of-speech
tagging \cite{Bengio2003NNLM}, semantic labeling and more recently, translation. The main interest
of such works is that they allow to take into account both long-range term dependencies
and semantic relatedness of words, thanks to the distributed representation of words in a vector space.
However, most of the works focused on building generic language model, i.e.
what would be called in IR a ``background language model''.
Such language models cannot be used directly for ad-hoc IR tasks, since, due to their huge number of parameters, learning accurate individual models for each document of the collection %  appears intractable %to %would be impossible
%to perform accurate learning of these models for learn the appropriate parameters
using maximum likelihood techniques
like for classical unigram multinomial language models appears completely intractable. Equally importantly,
the learned document model would be over-specific -- the hypothesis that the document language model generate the query would not be held anymore.

An interesting alternative was proposed by
Le and Mikolov \cite{Le2014DistributedDocuments} who recently published a neural network language model  in which they propose to represent a context (a document, or a paragraph) as a vector that \emph{modifies} the language model, which avoids building costly individual models for each document to consider.
Our work is based on the findings of Le and Mikolov \cite{Le2014DistributedDocuments}.
In this work, we follow a distributed approach where the document is represented
by a vector in $\real^n$. %Our work is directly based on this type of model and o
Our contributions are the following:
\begin{enumerate}
\item We generalize the model proposed in \cite{Le2014DistributedDocuments}, defining a more powerful architecture %model
and new ways to %take
consider individual specificities of the documents;
\item We apply the model for ad-hoc Information Retrieval;
%parameterize the language model;
\item We perform intensive experiments on standard IR collections (TREC-1 to 8) and analyze the
results.
\end{enumerate}

The outline of the paper is as follows. We first briefly overview the many related works (Section
\ref{sec:related-works}), before exposing the language models
(Section \ref{sec:model}). Finally, experiments are reported (Section \ref{sec:experiments}).

\section{Related works}

\label{sec:related-works}

This section discuss related works by first presenting those dealing with the
term dependencies and mismatch problems. We then introduce related works about Neural
Network Language Models (NNLM).

%\cite{Yang2010Language}

\subsection{Handling term dependencies}

Early approaches for handling term dependencies in IR  considered extensions of the bag of word representation of texts,  by including bi-grams to the vocabulary. Such an approach was taken by Fagan \cite{Fagan1987PhraseIndexing} for
vector space models, while the language model counterpart were proposed in the late of 90s \cite{Song1999GLM,Srikanth2002BitermLM,Gao2004DP} where the authors proposed to use a mixture of the bigram and unigram language models, the difference being in how to estimate the bigram language model or on how bigram are selected (\cite{Gao2004DP} used a dependency grammar parsing to generate candidate bigrams).
%handled bi-grams
% by using specific n-gram language models or defining some formal grammars to encode dependencies. In every case, the main problems are related to how handling sequential units %those new units,
% and whether terms in bi-grams should be included
This approach has proved to be not so successful,
most probably because more complex units imply sparser data \cite{Zhai:2008fg}, which in turn
implies inaccurate bigram probability estimations.
An alternative, where the mixture is defined within the quantum probabilistic framework, was proposed by
Sordoni et al.\cite{Sordoni2013term-dependencies}. This work proposed an elegant way to combine unigram and bigram distributions by leveraging this new probabilistic framework. In this paper, we investigate another principled way to model distributions over n-grams, with $n$ not being restricted to 1 or 2.

%Gao et al. \cite{Gao2004DP} uses dependency grammar to identify relevant sequences of terms.
More sophisticated probabilistic models have been proposed to deal with more terms, as well as different dependency constraints between terms.
Region-based proximity models combine the score of several models, each dealing with a specific dependency between query terms. For example, one of the submodels could be computing a score for three query terms co-occurring close together in the document.
Metzler and Croft \cite{Metzler2005MRF} proposed a Markov Random Field approach where each clique corresponds to a set or sequence of query terms. This work was extended by Bendersky et al. \cite{Bendersky2012HOT} who  considered sets of concepts (phrases or set of words) instead of set of words.
In a different probabilistic framework, Blanco \cite{Blanco2012BM25Operators} proposed to extend BM25F, a model able to take into account different source of evidence to compute the importance of a term for a document, to take into account term proximity by defining operators on so-called virtual regions.
In this work, these works are somehow orthogonal to ours since we are not interested by combining various sources of evidences, but rather by investigating whether a parametric language model can capture typical sequence of terms.

\subsection{Vocabulary mismatch}

One of the most used techniques to deal with the problem of vocabulary mismatch is query expansion, based on pseudo-relevance feedback, whereby terms are added to the query based on a set of (pseudo) relevant documents \cite{Manning2008IR}. It has been shown to improve search results in some cases, but is prone to the problem of query drift, which can be controlled using statistical theories like the portfolio theory \cite{CollinsThompson2009QueryExpansion}. Using pseudo-relevance feedback is orthogonal to our work, and could be used as an extension to estimate a query relevance language model \cite{Lavrenko2010Generative}.

Global analysis can be used to enrich the document representation by using co-occurrence information at the dataset level. For example, Goyal et al. \cite{Goyal2013Neighborhood} use a term association matrix to modify the term-document matrix and account for term relatedness. We believe that such information to be encoded in the neural language model we propose to use in this paper.

Dimensionality reduction techniques such as latent semantic models
have been proposed for dealing with vocabulary mismatch issues \cite{Deerwester:1990gu}. The idea
is to represent both the document and the query in a latent space,
and to compute a retrieval score based on these representations.
However,
such models do not work well in practice because many document specific
terms are discarded during the dimensionality reduction process \cite{Wang2013RLSI}.
It is thus necessary to combine scores from such latent models with scores from standard IR approaches such as BM25 to observe effectiveness improvements.  This approach has been followed by \cite{Wang2013RLSI} in vector spaces and Deveaud et al. \cite{Deveau2013QueryConcepts} for probabilistic (LDA) models. In the latter work, a latent-based language model is used as a background language model.
In this paper, we consider a combination of a document-specific neural network language model with a standard unigram multinomial one.

%In our work, we use a neural network-based language model.
\subsection{Neural Network Language Models}

The idea of using neural networks to build language models emerged in the last ten years. This area of research is included in the recent very active field of ``representation learning''.
Bengio et al. \cite{Bengio2003NNLM} is one of the first works that model text generation (at a word level)
using neural networks. The model does so by using a state (a vector in $\mathbb{R}^n$) to represent the history. This state defines a probability distribution over words, and can be updated with a new observation, hence allowing to define a language model over a text.

Such an idea of representation of texts in some latent space has been widely explored since then.
A recent successful work is the  well-known {\it Word2Vec} model \cite{Mikolov2013Word2vec} that proposed
a simple neural network architecture to predict a term within a predefined window. This model is fast to learn, thus allowing to compute distributed representations of words over large collections, and has also been shown to implicitly encode relationships (syntactic and semantic) between words: for example, there exists a translation that transforms the representation of a singular word (e.g. ``computer'') into the representation of its plural form (``computers''). Other works based on similar ideas were applied to sentiment detection \cite{Socher2012Compositionality} or automated translation \cite{Schwenk2013Translation}.

%  propose new representation of texts
%One of the first works  neural network language models is  %The idea of using representing texts
%

% of words in a continuous representation space, where their relative positions can be used to determine their dependencies. Using an embedding is a way to add a regularization over words representations : two words with similar meanings tend to be projected close to each other, and are thus likely be observed in a similar context. Such an  approach can thus be used to predict sequences of words, or the next word in a text knowing a previous sequence of words.
% The idea to use embeddings
%to predict sequences of elements was also developed in language models, such
%as in the case of the well-known Word2Vec model [21]. Once again, using an
%embedding is a way to add a regularization over words representations : two
%words with similar meanings tend to be projected close to each other, and will
%thus be predicted in a similar context.but been
Closer to IR, the idea of representing the history as a state, as in \cite{Bengio2003NNLM},
in a vectorial space has been exploited by Palangi et al. \cite{Palangi:2014vk} who proposed to
use the state obtained at the end of the document (resp. the query) as a vectorial representation of the whole
document (resp. the query). The relevance score is then equal to the cosine between the query and document vectors. They trained the model over clickthrough data and observed that their model
was able to better rank clicked documents above the others.

Compared to this work, our approach does not rely on external click data, and has the advantage
of conditioning the language model on the document vector, thus being less influenced by the
end of the document. It is based on the idea of using parameters to modify a generative
probabilistic model. This is what we call hereafter a parametrized model.

Parameterized probabilistic models have  been first applied to speaker
recognition. The need arise because those systems have to adapt fast
enough to a new user. Few researchers have tackled this problem by
designing a HMM whose probability distribution depends on contextual
variables. % (i.e. the context, that we note h).
Wilson and Bobick \cite{Wilson:1999dk}
proposed probabilistic models where the means of Gaussian distribution
vary linearly as a function of the context. As the output distribution
depends not only on the state but also on the context, a model may
express many distributions with a limited number of additional parameters.

This idea has been exploited by Le and Mikolov \cite{Le2014DistributedDocuments},
who proposed a parameterized language model which they experimented for sentiment
analysis and an information related task where relationships between query snippets are encoded by distances of their representations in the considered projection space. % closer representations need to set  % the distance between
%two
%for related query snippets %should be less
%than for unrelated ones.
We propose in this paper to extend this approach, by designing more sophisticated
models dedicated for ad-hoc IR tasks and experimenting them on various IR corpora (TREC-1 to 8). %with real information
%retrieval tasks  and experimenting more .

\section{Neural Network IR Models}

\label{sec:model}

In this section, we first present some background on classical language models for IR. Then, we present our contribution that allows the resulting IR model to deal with term dependencies and cope with term mismatch issues using representation learning techniques. At last, we present a parametric extension that performs document-dependent transformations of the model.

As most of the models deal with sequences, to clarify and shorten notations, we define $X_{i\ldots j}$ as the sequence $X_i, X_{i+1},\ldots,X_{j-1},X_j$ and suppose that  if $i>j$, the sequence is empty. %, and $f(X_{i\ldots j})$ as the sequence  $f(X_i),\ldots,f(X_j)$ .

\subsection{Background: Language Models for IR}

Language models are probabilistic generative models of text -- viewed as a sequence of terms.
If a text is composed of a sequence of terms  $t_{1\ldots n}$, where each $t_i$ corresponds to a word
in a pre-defined vocabulary, we can compute
the probability of observing this sequence given the language model $M$  as:
\[
P\left(t_{1 \ldots n} | M \right) = \prod_{i=1}^{N} P\left(t_{i}|t_{1\ldots i-1}, M\right)
\]
Language models are used in IR as a simple yet effective way to compute the relevance
of a document to a query \cite{Zhai:2008fg}. There exists different types of language models for IR,
but one of the most standard is to equate the relevance of the document $d$
with the likelihood of the language model generating the query, using the evaluated document language model $M_d$.
\[
P\left(d\mbox{ relevant to }q\right) = P\left(q\right | M_d)
\]where  $M_d$  is the so-called \textit{document} language model, which is the model within the family of models $\mathcal M$ that maximizes the probability of observing the document $d$ composed of terms  $d_{1 \ldots N}$, that is:

\begin{align}
\label{eq:lm-ml}
M_{d}&=\argmax_{M \in \mathcal M}P\left(d | M \right)\\\nonumber
			   &=\argmax_{M \in \mathcal M} \sum_{i=1}^{N} \log P\left(d_i|d_{1\ldots i-1}, M\right)
\end{align}

Among the different families of generative models, the n-gram multinomial family is the most used in IR, with $n$ usually equal to 1.
The multinomial model assumes the independence of terms given the $n-1$ previous ones in the sequence.  Formally, if  $M$ is within this family, then
\begin{eqnarray*}
P\left(t_{i}|t_{1\ldots i-1}, M\right) & = & P\left(t_{i}|t_{i-n+1 \ldots i-1}, M\right)  \\
-& = & \theta(t_i | t_{i-n+1 \ldots i-1})
\end{eqnarray*}
where  $\theta$  is a conditional probability table giving the probability of observing the term  $t_i$  after having observed the sequence $t_{i-n+1\ldots i-1}$.

For a given document, the parameters $\theta$  that maximize equation (\ref{eq:lm-ml}) can be computed in a closed form formula:
\begin{equation}
\label{counts}
 \theta(t_i | t_{i-n+1\ldots i-1}) =
 \frac
   {\mbox{Count of } (t_{i-n+1 \ldots i} ) \mbox{ in } d}
   {\mbox{Count of } (t_{i-n+1 \ldots i-1} \bullet) \mbox{ in } d}
\end{equation}
where $\bullet$ correspond to any term of the vocabulary (i.e. $t_{i-n+1 \ldots i-1} \bullet$ corresponds to the sequence $t_{i-n+1},\ldots,t_{i-1}, u$ where $u \in \mathcal{W}$).
With $n=1$, we get a simple unigram model that does not consider the context of the term (note that in that case the denominator is equal to the length of the document). For instance, in a document about Boston, ``trail'' is more likely to occur after ``freedom'' than in other documents. This information is then important to take into account to build more accurate IR models, since a \textit{good} document model about Boston would give a higher probability to ``trail'' occurring after ``freedom'', and thus the corresponding documents would get a higher score for queries containing the sequence ``freedom trail''.
Works like \cite{Song1999GLM} have  explored the use of language models with $n>1$. In such cases, the models are able to capture term dependencies but usually at the cost of higher complexity and loss of generalization due to the sparsity of the data -- longer sequences are more unlikely to occur in a document $d$, even for sequences that are strongly related with $d$ in terms of content from a user perspective. % correspond to %could \textit{they could have} in a hypothetical document, equivalent to $d$  in terms of content from a user perspective.
With $n\ge 2$, the estimated probabilities would be in most cases equal to 0.

Even with $n=1$, the estimation given by the maximum likelihood might be wrong, and discard documents just because they do not contain one query term, even if they contain several occurrences of all others.  % does not occur in the document under consideration.
To avoid such problems, smoothing techniques are used to avoid a zero probability (and thus a score of 0) by mixing the document language model with a collection language model\footnote{It can be the collection the document belongs to, or any external collection of documents} denoted $M_C$. A collection language model $M_c$  correspond to the language model that maximizes the probability of observing the sequences of terms contained in the document of the collection $C$.

A standard smoothing method is the Jelinek-Mercer one that consists in a mixture of the document language model and the collection language model. Formally, given a smoothing coefficient $\lambda \in [0,1]$, the language model of the document becomes:
\begin{eqnarray}
P(t_i | t_{i-n+1 \ldots i-1}, \lambda, d,C)
=   (1-\lambda) P(t_i | t_{i-n+1 \ldots i-1}, M_d)
+ \lambda P(t_i | t_{i-n+1 \ldots i-1}, M_C)
\label{eq:jm-smoothing}
\end{eqnarray}

Even for low values of $n$, the collection-based smoothing might not be effective.
In the following, we propose to develop new language models, that consider more sophisticated methodologies for smoothing,
able to  deal with long term dependencies and vocabulary mismatch issues.

\subsection{Neural Language Models for IR}
\label{sec:nnlm}

Distributed representations of words and documents have been known for long in IR as Latent Semantic Indexing techniques were introduced in 1999 \cite{Deerwester:1990gu}. They are useful since they overcome the sparsity problem we just evoked by relying on the spatial relatedness of the embedded objects. For example, the words ``cat'' and ``dogs'' should be closer together than ``cat'' and ``pencil'' in the vector space. This has been exploited in IR to deal with vocabulary mismatch problem, but the idea of leveraging this kind of representation for language models is more recent \cite{Bengio2003NNLM}. Such language models are built using neural networks (hence their name neural network language models) and offer several advantages:
\begin{itemize}
\item Compression abilities offered by representation learning techniques allow us to consider longer term dependencies (i.e., longer n-grams) than with classical approaches; % in the language model);
\item Geometric constraints implied by the continuous space used to represent words induce some natural smoothing on the extracted relationships, which enables better generalization abilities and avoids well-known difficulties related to zero counts of query words (or n-grams) in the considered documents. %This also allows to
\end{itemize}

In this work, we propose to include such a distributed language model in the classical probability computations of IR query terms. Thus, rather than equation \ref{eq:jm-smoothing}, we propose to consider the following probability computation:
%to replace the collection model  $M_C$  in equation (\ref{eq:jm-smoothing}) by
\begin{eqnarray}
P(t_i | t_{i-n+1 \ldots i-1}, \lambda,\gamma, d,C)
=   (1 - \lambda) ( (1-\gamma) P(t_i | M_d)
+ \gamma P(t_i | M_C))  + \lambda P_{NN}(t_i| t_{i-n+1 \ldots i-1}, d, C)
\label{eq:nn-smoothing}
\end{eqnarray}
where $P_{NN}(t_i| t_{i-n+1 \ldots i-1}, d, C)$ is the probability of observing the term $t_i$ after the sequence $t_{i-n+1 \ldots i-1}$ in the document $d$ of the collection $C$ according to our neural network model. It corresponds to introduce term dependencies and vocabulary proximity in a classical unigram language model that is not able to capture such relationships.

Two version of our model are detailed hereafter:
\begin{itemize}
\item A \textit{generic} neural network language model defined by a background collection (section \ref{sec:gen-nnlm}) ;
\item A \textit{document-specific} neural network language model estimated by a background collection and the document at hand (section \ref{sec:doc-nnlm}). Note that in that case, we are interested by the performance of the model when $\lambda$  is close to 1 (ideally, 1) since this would mean that the document-specific model is specific enough to fully represent the document.
\end{itemize}

% where the first term corresponds to a classical unigram language model learned following formula \ref{counts} and the second corresponds to our neural model, which can  either be defined as a generic model for smoothing or as a document-dependent one that also involves individual term dependencies for each document.

Both generic and document-dependent models are represented in figure \ref{fig:model}, where the black part corresponds to the common framework for both models %, the green part corresponds to the distributions computation for the generic model
and the green and blue parts respectively stand for specific layers for the generic and the document-dependent models.  %the blue part %includes %
%stands for
%an additional layer included in the document-dependent model. % the  illustrates the neural network model that we use in this paper.

\subsubsection{Generic Neural Model}

\label{sec:gen-nnlm}
% Classical language models determine individual probabilities for each sequence of $n$ terms, which quickly becomes intractable when $n$ increases. In practice, only unigrams, bigrams or tri-grams are considered.

There are two types of neural network language model, those that take into account an infinite context, and those that only take into account a limited number of previous terms. The former are based on recursive neural networks, while the latter are standard feedforward ones. In this work, we investigate the use of the feedforward networks since they are easier to train. Moreover, we expect that the document-specific model that we describe in the next section captures longer term dependencies.

The input of the neural network corresponds to the $n-1$ previous terms. For the first $n-1$ words of a document, we use a special ``padding'' term. To each term  $t_i$  corresponds a vector $z^t_i$ in the vector space $\mathbb{R}^{m_0}$. The  $n-1$  vectors are then transformed through a  function $\phi$ (descr below) into a state vector $s$  in $\mathbb{R}^{m_f}$ where $f$  is the index of the last layer.   This state vector purpose is to summarize the contextual information, which is then taken as an input by the last layer (HSM in the figure) that computes a probability for each term of the vocabulary.

% Representation learning techniques that recently emerged offer the possibility of considering longer term dependencies by using distributed representation of words in a vector space  $\mathbb{R}^k$.
% The representation of a word which the locations  $z_w$  of a words are used to compute conditional probabilities used in the language model.

% In that way, the conditional probability of a given word $t_i$ %, knowing a past sequence of query words $(q_{i-n+1} \ldots  q_{i-1})$,
% depends on the representation of the past sequence of query words $t_{i-n+1, \ldots, i-1}$, %its components
% by considering $\phi(z^q_{i-n+1, \ldots,  i-1})$,
%\begin{equation}
%\label{probaQi}
%P_{Neural}\left(q_{i}|(q_{i-n+1} \ldots q_{i-1})) \propto \Phi(q_i,\phi(z_{q_{i-n+1}} \ldots  z_{q_{i-1}}))
%\end{equation}
% where $\phi:\mathcal{Z}^{n-1} \rightarrow \mathbb{R}^k$ maps a summary vector from the concatenation of $n-1$ representation vectors in $\mathcal Z$. % of  words. % and \Phi(q_i,

In our experiments, we considered three different functions for  $\phi$, each of which being a composition of different functions. Following the neural network literature, we term each of the component functions a layer.

\paragraph{Model 1 (M1):  $\textrm{linear}(m_1)-\tanh$}
The first layer   transforms linearly the  $n-1$  vectors $z_j\in\mathbb{R}^{m_0}$  in  a vector in $\mathbb{R}^{m_1}$, that is:
$$
l_1 = \sum_{j=1}^{n-1} A_j z_{j} + b
$$
where each  $A_j$ is a matrix of dimension $m_1 \times k$ and $b$ is a bias vector in $\mathbb{R}^{m_1}$.% that corresponds to a bias term that avoids null inputs for the HSM layer.
The second layer introduces a non-linearity by computing the hyperbolic tangent ($\tanh$)  of each of its inputs
$$
\forall j \ l_{2j} = \tanh(l_1j)
$$
In this model, the function $\phi$ has  $(n-1) \times m_0 \times m_1$  parameters.

\paragraph{Model 2 (M2):  $\textrm{linear}(m_1)-\tanh-\textrm{linear}(m_2)-\tanh$}
The second function $\phi$  we consider is an extension of the first one where we add a linear (matrix $B$ of dimension $m_2 \times m_1$) and a non-linear layer ($\tanh$). In this model, the function $\phi$ has $(n-1) \times m_0 \times m_1 + m_1 \times m_2$  parameters.

\paragraph{Model 3 (M2Max):  $\textrm{linear}(\kappa m_1)-\textrm{max}(\kappa)-\textrm{linear}(m_2) -\tanh$}
In the third model, we substitute to the second layer (hyperbolic tangent) of the previous model another non-linear function, a maximum pooling layer. Maximum pooling layers are useful in deep neural networks because they introduce an invariant \cite{Kalchbrenner2014ConvolutionSentiment}, i.e. they allow to learn more easily that ``big'' in a ``a big blue cone'' and ``a big cone'' has the same influence on the next term to appear. It is defined by a parameter $\kappa$ (set to 4 in our experiments)
$$
\textrm{max-pooling}_\kappa(x)_j = \max \{ x_{\kappa\times(j-1)}, \ldots,  x_{\kappa\times j - 1} \}
$$ In this model, the function $\phi$ has  $(n-1) \times \kappa \times m_0 \times m_1 + m_1 \times m_2$  parameters.

Then, from a sequence of $n-1$ terms $(t_{1\ldots n-1})$, we get a summary vector  $\phi(z_{1\ldots  n-1})$ that is used to compute probability distributions over terms -- the probability that each term in the vocabulary occurs after the sequence of $n-1$  terms.

In our model, we use a Hierarchical SoftMax (HSM),  which corresponds to an efficient multivariate differentiable function that maps an input vector (given by $\phi$ ) to a vector in $\mathbb{R}^\mathcal{V}$  whose values sum to 1 \cite{Morin:2005vo}. It allows us to compute the probability $HSM_t(v)$ of a term $t$ given an input vector $v$, with  a complexity logarithmic with respect to the number of words in the vocabulary. The HSM is associated to a (binary) tree where leaves correspond to words. Formally, the   function is defined as: %:
\begin{equation}
\label{HSM}
HSM_t(v)=\prod_{s \in \textrm{path}(t)} \frac{1}{1 + \exp(b_s(t) \times x_s \cdot v)}
\end{equation}
where $HSM_t$ denotes the component corresponding to the word $t$ in the distribution encoded by the $HSM$ layer, $v$ corresponds to the input vector given to the function, $\textrm{path}(t)$ corresponds to the set of nodes leading to the leaf corresponding to word $t$,  $x_s$ is the vector associated to the inner node $s$ of the tree, and $b_s(t)$  is -1 (resp. 1) if the word  $t$  can be accessed through the left (resp. right) branch from node $s$. In our case, the tree of the HSM function is a Huffman tree Starting with trees composed of one node (the terms) associated with a weight (the number of occurrences in the collection), the algorithm combines iteratively the two trees with the lowest weights into a new tree formed by a new node with two branches leading to the two selected trees. The set  of vectors $x_s$ associated to each node $s$ correspond are parameters of the model.

This allows us to easily compute  a distribution of conditional probabilities $P_{NN}(t|t_{1 \ldots n-1})$ for the next word $w \in \Omega$ knowing the past sequence of  $n-1$ observed words: %  following conditional probability computation for each word $w \in \Omega$:
\begin{equation}
\label{probaWi}
P_{NN}(t|t_{1\ldots n-1})=HSM_t(\phi(t_{1 \ldots n-1}))
\end{equation}

At last, to get an operational model that enables the computation of the probability of a given query with regards to the generic model, one has to learn the representation of words, the parameters of the HSM and the parameters of the functions. We denote this set of parameters by $\Theta$. The learning problem can then be formulated as the maximization of the probability of observing the documents $d$ in the collection $\mathcal D$ :
\begin{equation}
\Theta^*=
\argmax\limits_{\Theta}
\sum\limits_{d \in {\mathcal D}}
\sum\limits_{i=0}^{|d|} w(d_i) \log  \ HSM_{d_i}(\phi(z^d_{i-n+1 \ldots i-1}))
\label{eq:cost}
\end{equation}
where $w(d_i)$  is a weight used to lower the importance of  frequent terms. This was used by Mikolov \cite{Mikolov2013Word2vec} to ensure a better learning, and we used the same settings.
\note{Donner la formule de w}

\subsubsection{Document-Dependent Model}

\label{sec:doc-nnlm}

\begin{figure}
\begin{center}

\input{fig-model}

\end{center}
\caption{General architecture of the generic (black and green/dashed) and document-dependent (black and blue/dotted) language models }
\label{fig:model}
\end{figure}
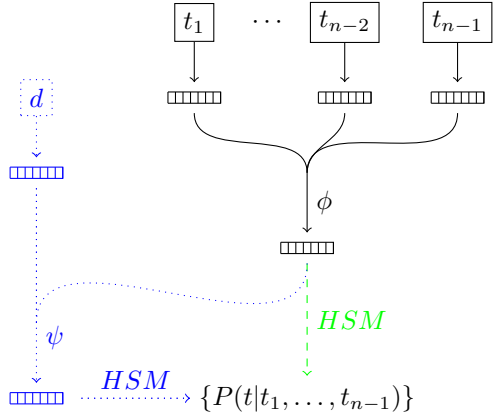

The generic neural network language model presented above handles long term dependencies and semantic relatedness between words  for all documents in the collection. This language model can be a good alternative to the multinomial unigram collection language model used for smoothing.
However, we believe that taking into account specificities of the document at hand leads to a better language model, and hence to better retrieval results. Indeed, as explained above on an example about Boston, term relationships can be different from documents to others.
%Similarly, two terms strongly semantically related in some context can have a significantly different meaning in other ones. For instances,

Learning specific neural language models for all individual documents is unfeasible for the same reasons as for n-gram language models for $n>1$: Learning  term dependencies and semantic relatedness on the sequences contained in a single considered document is likely to lead to over-training issues due to the lack of vocabulary diversity and sequence samples.
To overcome this problem, we follow the approach  of \cite{Le2014DistributedDocuments} where the probability distribution of a generic neural network language model is modified by a relatively small set of parameters that can be reliably learned from the observation of a single document: The dimension of such a vector (100-200) is typically much smaller than the number of parameters of a multinomial distribution (size of the vocabulary).
Following this approach is interesting since it allows to first learn a language model from the whole collection, and then only learn \textit{how to modify it} for a specific piece of content. Using parameters to modify the behavior of a generative probabilistic model has been used in many works in signal processing, like gesture recognition \cite{Wilson:1999dk}, where the model has to be quickly adapted to a specific user: Such models benefit from a large source of information (the collection of all gestures or documents in our case) and at the same time can be made specific enough to describe an individual user or document.

The modification of the neural network language model is shown in figure \ref{fig:model} (blue/dotted part). A document-specific vector $z_d\in \mathbb{R}^{m_f}$ is used to modify the generic language model learned on the whole collection, by modifying the vector representing the context of the word to appear. The modified vector is then used to generate a probability distribution by using, as for the generic language model, a HSM.

% Then, while the embeddings of words are common for the whole collection in concern, a specific representation vector $z_d \in {\mathbb R}^k$ is determined for each document $d \in {\mathcal D}$. In our document-dependent neural model, the HSM function determining probability distributions of terms is applied on the result of a modified state vector  resulting from a ``merging'' operation of the generic state vector $s$ obtained from $\phi$ with the document representation $z_d$.

In this paper, we consider two ``merging'' operations $\psi:{\mathbb R}^{m_f}\times {\mathbb R}^{m_f} \rightarrow {\mathbb R}^{m_f}$ that associate the state vector $s$  given by $\phi$  and the document specific vector $z_d$  to
\begin{itemize}
\item their sum, i.e. $\psi(s,z_d)=s+z_d$
\item their component-wise product, i.e. $\psi(s,z_d)=s\odot z_d$
\end{itemize}
These two functions are simple yet they can substantially modify the distribution of the language model. Taking again the example of Boston, the components of  $z_d$  would bias the model to words likely to occur in such documents, thereby e.g. increasing the probability to find ``trail'' after ``freedom''. This is done by making the state vector more orthogonal to vectors in the HSM that lead to ``trail'' than to other words associated to ``freedom''.

Note that, because the solution of our optimization problem is not unique, equivalent solutions (in term of the cost function) could be obtained by rotation -- this would in turn have an impact on the benefit of using such modifications. Our experiments show however that there is a gain associated to using even simple functions like term-wise multiplication or addition. Future works will explore more sophisticated and appropriate transformations of the state vector.

In theory, this document-specific language model could be used alone (i.e. without any smoothing, since it is based on a background LM) to estimate the likelihood of generating a query. In practice, as shown in the experiments below, the model is not yet powerful enough to be used so. However, combined with a classical multinomial unigram model as proposed by equation \ref{eq:nn-smoothing}, it allows to observe interesting improvements for IR ad-hoc tasks. This corresponds to a first step towards a principled solution for handling vocabulary mismatch and term dependencies.

\section{Experiments}

\label{sec:experiments}

We used the TREC-1 to 8 collections for experimenting our models. We used as a baseline BM25 \cite{Robertson2009The-Probabilistic-Relevance} with standard parameter settings ($k_1=1.2$ and $b=0.5$) since it has a reasonable performance on these collections. We left out the comparison with stronger baselines since we were interested first in comparing the different models between themselves. The collection was pre-processed using the Porter stemmer, with no stop words.

To learn our language models, we pre-processed the collection using the same pre-processing (no stop words, Porter stemmer), but removed words occurring less than 5 times in the dataset since their learned representation would have been wrongly estimated. The vocabulary size we obtained was of 375,219 words. We used the word2vec \cite{Mikolov2013Word2vec} implementation\footnote{Code available \url{https://code.google.com/p/word2vec/}}  to pre-compute the initial word representations and hierarchical SoftMax parameters. We used standard word2vec parameters (window of size 5, and used an initial learning rate of 0.5) and iterated 5 times over the corpus, thus ensuring that the initial representation of words is meaningful and easing the following learning process.

\begin{table*}
\begin{center}
\begin{tabular}{|c|c|c|c|}
\hline
% Number of terms = 375,219
Name & Model &  $\phi$ &  word/HSM \\ \hline
M1 & linear(100) - tanh & 40,000 & 75,043,700 \\ \hline
M2 & linear(100) - tanh - linear(100) - tanh & 50,000 & 75,043,700  \\ \hline
M2Max & linear(400) - max(4) - linear(100) - tanh & 170,000 & 75,043,700  \\ \hline
\end{tabular}
\end{center}
\caption{Models and number of parameters (resp. for the function $\phi$, the term representation and HSM classifier, and in total) }
\label{tab:nparams}
\end{table*}

Then, we first trained the generic language models (Section \ref{sec:gen-nnlm}) that are listed in table \ref{tab:nparams} along with the number of parameters, where we distinguish the computation of the context vector ($\Phi$) and the parameters representing words (initial representation and HSM parameters). We can see that most of the parameters are related to words.

% # of documents in all TRECs = 1634243
% at least 5 occurrences (supposing in different documents)
% 500000 documents seen
% => 1.52976 views !

Following common practice with stochastic gradient descent,
we used a steepest gradient descent to learn our terms representation, using the following decreasing learning rate:
$$
\epsilon_{k} = \frac{\epsilon_0}{1 + k \times \delta }
$$where $\epsilon_0=0.1$ and $\delta=2e-4$. Following the literature in neural language model learning, We used mini-batch stochastic optimization with batches of 100 documents. After 50000 iterations, the average perplexity was around 2. %\note{Dire ce a quoi correspond une iteration : 100 docs c'est ca ?}
The number of iterations has thus been set empirically to 50000, as it corresponded to (1) having seen
(in expectation) at least 5 times each word of the collection and
(2) the likelihood was not evolving anymore.

For each topic, we selected the top-100 documents as ranked by BM25, and learned the  document parameters $z_d$. We also used a gradient descent\footnote{We used a different technique here since we wanted a fast convergence over a small set of parameters and a single document} with resilient backpropagation (Rprop) \cite{Riedmiller1993RProp} using an initial learning rate of $0.1$, and iterated until the difference between two successive settings be under $1e-4$.

\begin{figure*}
\begin{center}
\includegraphics[width=1\textwidth, height=0.8\textheight, keepaspectratio=true]{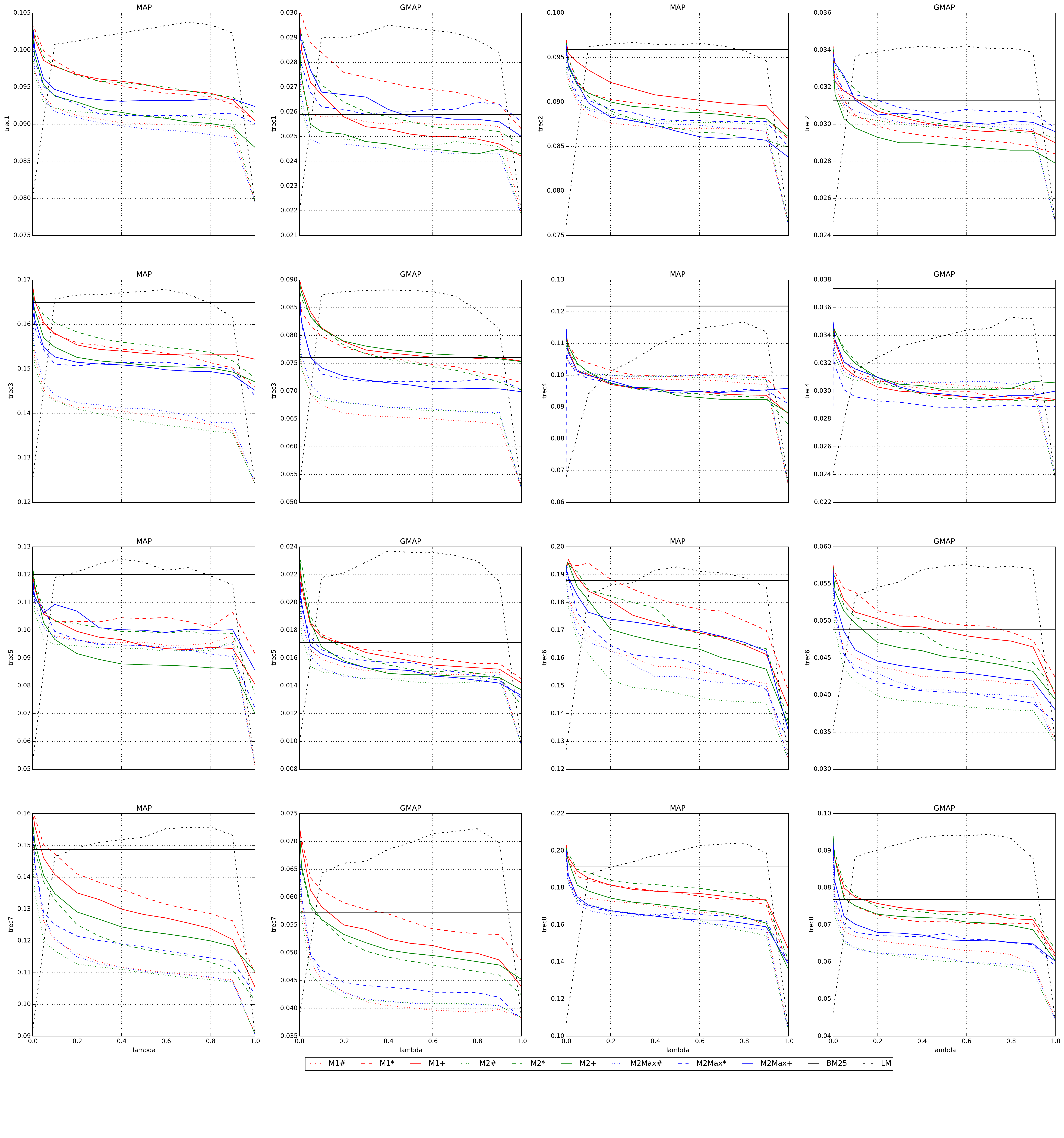}
\end{center}
\caption{
MAP and GMAP for various TREC datasets for BM25, the Jelinek-Mercer model (LM) and our neural network models (the generic language model is specified by a \#).
The x-axis is the value of $\lambda$ (in equation \ref{eq:jm-smoothing} for LM and in equation \ref{eq:nn-smoothing} for our models). We set $\gamma$ to 0.5 in Equation \ref{eq:nn-smoothing}
}

\label{fig:performance}
\end{figure*}

Results are reported in Figure \ref{fig:performance} and table \ref{fig:performance2} where we compare the different IR models  for MAP and GMAP metrics on all TREC datasets. In this figure $LM$ stands for the classical unigram multinomial language model with Jelinek-Mercer smoothing (equation \ref{eq:jm-smoothing}). Our 9 models are noted by the name of the model (M1, M2 or M2max following which $\phi$ function is used) followed by the operator used to transform the representation with the document vector (\# for none which corresponds to the generic model, $+$ for the addition operator and $*$ for the component-wise multiplication). %  , generic and document-specific neural language models).
The x-axis is the value of $\lambda$ (in equation \ref{eq:jm-smoothing} for LM and in equation \ref{eq:nn-smoothing} for our neural models). We set $\gamma$ to 0.5 in Equation \ref{eq:nn-smoothing}.

%Results are reported in figure \ref{fig:performance} and table \ref{fig:performance2} for MAP and GMAP metrics on all TREC datasets, for the document-specific model.
We first observe that compared to classical Jelinek-Mercer LM, there is a slight overall improvements with our document-specific models for both metrics. Furthermore, there is greater stability with respect to the smoothing coefficient $\lambda$, where the best values are for small values of $\lambda$. We also note that the best improvements are for badly performing topics (since GMAP is more improved than MAP).

Comparing the different models, we see that the generic language models perform worse than their specific counterparts. In table \ref{fig:delta_z}, we compared the performance of the model when using a generic and a document-specific language model (we set $\lambda$ to 0.01), and observed that in the vast majority of cases we improved the performance by using a document specific model.

While there is a difference on each TREC dataset, there is no clear pattern between the different models M1, M2 and M2Max. It seems however that more complex models do not perform better. We hypothesize that there is more to gain when trying different merging functions $\psi$.

We also observe that taking $\lambda=1$ for our  %for extreme values of $\lamda$, the
 document-specific language models, which corresponds to only considering probabilities from our neural models without any other form of smoothing, does not degrade as much as extreme values of $\lambda$ for the %document and collection
classical multinomial unigram model LM  ($\lambda=0$ for the document model only or $\lambda=1$ for the collection one).
This is an encouraging result since it shows that using document specific parameters enables to well capture some specificities of the documents. It actually modifies the term representation, leading to better results than a generic language model.

Document length has also an influence on the latter observation: We observe that for TREC-1 to TREC-4 the drop is lesser than for TREC-5 to TREC-8. This is to be related to the average length of documents which is higher for latter TREC datasets. It shows that our document specific models are less effective for longer documents, and that more sophisticated models for modifying the context state with the document specific parameters are needed.

\begin{table*}
\begin{centering}
\small
\begin{tabular}{|l|cccccccc|cccccccc|}
\hline
{}  & \multicolumn{8}{c|}{MAP } & \multicolumn{8}{c|}{GMAP } \\
\cline{2-17}
{}  & 1  & 2  & 3  & 4  & 5  & 6  & 7  & 8  & 1  & 2  & 3  & 4  & 5  & 6  & 7  & 8 \\
\hline
BM25   & \textbf{0.1} & \textbf{0.1} &          0.16 & \textbf{0.12} & \textbf{0.12} &         0.19 &          0.15 &         0.19 & \textbf{0.03} & \textbf{0.03} &          0.08 & \textbf{0.04} & \textbf{0.02} &          0.05 &          0.06 &          0.08 \\
LM     & \textbf{0.1} & \textbf{0.1} & \textbf{0.17} &          0.11 & \textbf{0.12} &         0.19 &          0.15 & \textbf{0.2} & \textbf{0.03} & \textbf{0.03} & \textbf{0.09} &          0.03 & \textbf{0.02} & \textbf{0.06} & \textbf{0.07} & \textbf{0.09} \\
M1\#    & \textbf{0.1} &         0.09 &          0.15 &           0.1 &          0.11 &         0.18 &          0.14 &         0.19 & \textbf{0.03} & \textbf{0.03} &          0.07 &          0.03 & \textbf{0.02} &          0.05 &          0.06 &          0.08 \\
M1*    & \textbf{0.1} & \textbf{0.1} &          0.16 &          0.11 & \textbf{0.12} &         0.19 & \textbf{0.16} & \textbf{0.2} & \textbf{0.03} & \textbf{0.03} &          0.08 &          0.03 & \textbf{0.02} & \textbf{0.06} & \textbf{0.07} & \textbf{0.09} \\
M1+    & \textbf{0.1} & \textbf{0.1} & \textbf{0.17} &          0.11 &          0.11 & \textbf{0.2} & \textbf{0.16} & \textbf{0.2} & \textbf{0.03} & \textbf{0.03} & \textbf{0.09} &          0.03 & \textbf{0.02} & \textbf{0.06} & \textbf{0.07} & \textbf{0.09} \\
M2\#    & \textbf{0.1} &         0.09 &          0.15 &           0.1 &          0.11 &         0.18 &          0.14 &         0.18 & \textbf{0.03} & \textbf{0.03} &          0.07 &          0.03 & \textbf{0.02} &          0.05 &          0.06 &          0.07 \\
M2*    & \textbf{0.1} &         0.09 & \textbf{0.17} &          0.11 & \textbf{0.12} &         0.19 &          0.15 & \textbf{0.2} & \textbf{0.03} & \textbf{0.03} & \textbf{0.09} &          0.03 & \textbf{0.02} & \textbf{0.06} &          0.06 & \textbf{0.09} \\
M2+    & \textbf{0.1} &         0.09 &          0.16 &          0.11 & \textbf{0.12} &         0.19 &          0.15 &         0.19 & \textbf{0.03} & \textbf{0.03} & \textbf{0.09} &          0.03 & \textbf{0.02} &          0.05 & \textbf{0.07} & \textbf{0.09} \\
M2Max\# & \textbf{0.1} &         0.09 &          0.15 &           0.1 &          0.11 &         0.18 &          0.14 &         0.18 & \textbf{0.03} & \textbf{0.03} &          0.08 &          0.03 & \textbf{0.02} &          0.05 &          0.06 &          0.07 \\
M2Max* & \textbf{0.1} &         0.09 &          0.16 &           0.1 &          0.11 &         0.19 &          0.14 &         0.18 & \textbf{0.03} & \textbf{0.03} &          0.08 &          0.03 & \textbf{0.02} &          0.05 &          0.06 &          0.08 \\
M2Max+ & \textbf{0.1} &         0.09 &          0.16 &          0.11 &          0.11 &         0.19 &  \textbf{nan} &         0.19 & \textbf{0.03} & \textbf{0.03} &          0.08 &          0.03 & \textbf{0.02} &          0.05 &  \textbf{nan} &          0.08 \\
\hline

\end{tabular}
\par\end{centering}

\protect\caption{MAP and GMAP for models trained for 50000 iterations with $\lambda=0.01$
and a modifying vector $z_{d}$ for various TREC datasets}
\label{fig:performance2}
\end{table*}

\begin{table*}
\begin{center}
\begin{tabular}{|l|rrrrrrrr|}
\hline
{} &       1 &       2 &       3 &       4 &       5 &       6 &       7 &       8 \\
\hline
M1*    &  0.0056 &  0.0024 &  0.0105 &  0.0061 &  0.0055 &  0.0119 &  0.0148 &  0.0093 \\
M1+    &  0.0044 &  0.0029 &  0.0127 &  0.0043 &  0.0046 &  0.0133 &  0.0120 &  0.0108 \\
M2*    &  0.0054 &  0.0028 &  0.0147 &  0.0035 &  0.0115 &  0.0149 &  0.0132 &  0.0167 \\
M2+    &  0.0026 &  0.0020 &  0.0125 &  0.0050 &  0.0096 &  0.0154 &  0.0156 &  0.0125 \\
M2Max* &  0.0012 &  0.0007 &  0.0049 & -0.0003 &  0.0005 &  0.0069 &  0.0010 &  0.0011 \\
M2Max+ &  0.0028 &  0.0013 &  0.0068 &  0.0024 &  0.0018 &  0.0071 &  0.0026 &  0.0038 \\
\hline
\end{tabular}
\end{center}
\caption{Average difference between the AP of the document-specific and generic models with $\lambda=0.01$.}
\label{fig:delta_z}
\end{table*}

% \begin{figure*}
% \begin{center}
% \includegraphics[width=.8\textwidth, keepaspectratio=true]{learning-cost}
% \end{center}
% \caption{Learning costs}
% \label{fig:learning-cost}
% \end{figure*}

% In figure \ref{fig:learning-cost}, we report the cost (equation \textbf{XXXX}) of the different models.

% \subsection{Query-wise performance}

% As with any new model for IR, it is interesting to look at the topics where the model increased, or decreased, the performance. In table \textbf{XXXX}, we report the difference between the average precision obtained by BM25 and the average precision of the model \textbf{XXXX} for the \textbf{XXX} biggest and lowest differences.

\section{Conclusion}

In this paper, we have proposed new %three
parametric %(document specific) and generic (collection specific)
neural network language models specifically designed for %and experimented on
adhoc IR tasks. We proposed various variants for both a generic and a document specific version of the model. %three versions of the
While the generic version of the model learns term dependencies and accurate representations of words at a collection level, %In the case of
our document-specific language model is modified for each document, through the use of a vector of small dimension, % and using them to  , which are which   %  this generic words representation for each considered document  by  introducing cluding a representation of the considered document in
%each individual document considering , the generic language model is modified through a vector that represents a document. This vector has
which only contains a few hundred parameters, and thus can be learned efficiently and estimated correctly from a single document. This vector is used to modify the state vector representing the context of the word for which we want to compute the probability of occurring, by using a component-wise multiplication or an addition operator.

We experimented with TREC-1 to TREC-8 IR test collections, where the top 100 results retrieved by BM25 were reranked using our generic and document specific language models. %, as well as with a standard LM with Jelinek-Mercer smoothing.
The results show that using a document-specific language model improves results over a baseline classical Jelinek-Mercer language model. We have also shown that the document-specific model obtained better results than the generic one, thus validating the approach document-dependent parameterization of the term representations. % where a generic language model is parameterized according to the document in concern.

While the results do not show a substantial improvement, we believe such models are interesting because, by improving the way a generic language model can be modified using a small set of parameters that represent a textual object (document, paragraph, sentence, etc.), we could finally reach a point where the document-specific model can make a big difference.
Future work will thus study different architectures  of the neural language model (including recurrent to consider longer dependencies), as well as use a relevance language model \cite{Lavrenko2010Generative} based on pseudo-relevance feedback -- this might be more reliable since language models will be learned from documents and applied on documents (rather than queries).

\bibliographystyle{abbrv}
\bibliography{ir-nnlm-ictir15}

\end{document}

%% file: fig-model.tex
\newcommand{\gnnvector}[1][black]{
   \begin{tikzpicture} \begin{scope}[scale=0.1]
    \foreach \i in {0,1,...,7}  { \draw[#1] (\i,0) -- (\i,1.5); };
    \draw[#1] (0, 0) -- (7, 0) -- (7, 1.5) -- (0, 1.5);
    \end{scope} 
    \end{tikzpicture}
}

\newsavebox{\nnvector}
\newsavebox{\bnnvector}

\savebox{\nnvector}{\gnnvector}  
\savebox{\bnnvector}{\gnnvector[blue]}

\begin{tikzpicture}

% Words
\node[rectangle,draw] (wd) at (0,0) {$t_{1}$};
\node[] at (1,0) {$\cdots$};
\node[rectangle,draw] (w2) at (2,0) {$t_{n-2}$};
\node[rectangle,draw] (w1) at (3.5,0) {$t_{n-1}$};

% Vectors
\node (xd) at (0, -1) {\usebox{\nnvector}};
\node (x2) at (2, -1) {\usebox{\nnvector}};
\node (x1) at (3.5, -1) {\usebox{\nnvector}};

\draw[->] (wd) -- (xd);
\draw[->] (w2) -- (x2);
\draw[->] (w1) -- (x1);

% Function f
\node (fx) at (1.5, -3) {\usebox{\nnvector}};
\coordinate (fxj) at (1.5, -2); 
\foreach \n in {xd, x2, x1} \path (\n) edge[out=-90, in=90] (fxj);
\draw[->] (fxj) -- (fx) node [midway,right] {$\phi$};

% HSM
\node[rectangle] (pw) at (1.5, -5) {$\left\{ P(t | t_{1}, \ldots, t_{n-1}) \right\}$};
\draw[green,dashed,->] (fx) -- (pw) node[midway, right] {$HSM$};

% Context
\node[blue, dotted, rectangle,draw] (d) at ([xshift=-3cm, yshift=2cm]fx.west){$d$}; 
\node[blue, rectangle] (z) at ([xshift=-3cm, yshift=1cm]fx.west){\usebox{\bnnvector}}; 
  %{ \begin{tabular}{c} $d$ \\ \usebox{\bnnvector} \end{tabular}};
\draw[blue,dotted,->] (d) -- (z);
\coordinate (mz_) at ([yshift=-2cm] z);
\node (mz) at ([yshift=-1cm]mz_) {$\usebox{\bnnvector}$};

\path[blue,dotted,draw] (z) edge (mz_) edge[->] (mz);
\path (mz_) -- (mz) node[midway, yshift=+0.2cm, right, blue] {$\psi$};
\path[blue,dotted] (fx) edge[out=-90, in=90] (mz_);
\draw[blue,dotted] (mz) edge[->] (pw);
\draw[blue,dotted] (mz) -- (pw) node[midway,above] {$HSM$};

\end{tikzpicture}